\documentclass[aps,tightenlines,twocolumn,superscriptaddress]{revtex4-1}
\usepackage{graphicx}
\usepackage{dcolumn}
\usepackage{bm}
\usepackage{csquotes}
\usepackage{epigraph}
\usepackage{amssymb}
\usepackage{color,graphicx}
\usepackage{subfigure}
\usepackage{lmodern}
\usepackage{amsmath}
\usepackage{amsbsy}
\usepackage{amsthm}
\usepackage{float}
\usepackage{bbm}
\usepackage{tikz}
\usetikzlibrary{quantikz}

\newcommand\tr{\operatorname{tr}}
\newcommand\ad{\operatorname{ad}} 
\newcommand\op{\operatorname{op}}

\begin{document}

\title{Can thermal quantum Gibbs states approve as quantum equilibrium states?}

\author{Ali Soltanmanesh}
\email[]{a.soltanmanesh@gmail.com}
\affiliation{Independent Researcher, Tehran, Iran}
\affiliation{Alumna, Department of Chemistry, Sharif University of Technology}
\begin{abstract}
\section*{Abstract}
In this study, we investigate a quantum harmonic oscillator interacting with a thermal bath of oscillatory fields in a quantum circuit. By solving the Lindblad master equation, we calculate the resulting interference pattern from measuring the system in the momentum space. Interestingly, we show that even if one considers the decoherence effect, the system will keep some of its quantum properties. Indeed, the equilibration does not completely leave the system in a Gibbs state, and the system remains coherent. Moreover we discuss the requirements of a process that can be called a thermalization. We show that in our system, the quantum thermodynamic equilibration process cannot be considered a thermalization. Also, we discussed that a Lindblad system-bath interaction cannot be explained by a thermalization process. Such an effect strongly can be detected when the frequency of the central system is high and the temperature is low.  Then, by introducing an entropy measure, we show that although the system is in maximum entropy, the equilibrium state is far from the Gibbs state.
\end{abstract}

\maketitle
\section{Introduction}

The coherence of a quantum system disappears in an interaction with a thermal bath. \cite{Joos,Weiss,Gold,Liu}. This problem is one of the main obstacles in developing quantum technologies \cite{DiV,Unr,Shor,Palm}. Also, decoherence is known as the main reason for quantum to classical transition \cite{Bre,Schl}.  There are a variety of schemes proposed to control the decoherence process in a system-environment interaction \cite{Gane,Cui,Rosz}. Nonetheless, the priority towards the solution to this problem is understanding the fundamental aspects that are responsible. Whenever a problem arises in the borderlines of physics, thermodynamics lies in the deepest layers of the problem. Nowadays, studying the evolution of quantum systems through the laws of thermodynamics is quite popular \cite{Gem,Gool,Vin,Jar,Croo,Pop,del,Bra}.

The laws of equilibrium thermodynamics can be applied to classical closed systems and to open subsystems. There is a belief that these laws are also applicable to open quantum subsystems \cite{Nie}. The decoherence theory makes it possible to study the thermodynamics of open quantum systems. Moreover one of the interesting aspects of thermodynamics is its description of changes of a state through decoherence. Many works with various perspectives have been done to study and define thermodynamic quantities for the open quantum systems by defining work, heat, and thermodynamic laws in the quantum regime. This includes thermodynamics of discrete quantum processes \cite{And}, the fluctuation theorems \cite{All,Esp,Tal}, defining the passivity condition of the equilibrium state for the general quantum systems \cite{Pus} and mathematical characterizations of these notions \cite{Hen,Wei,Deu,Esp2,Esp3,Gem2,Wil,San}.

Although many works have been done to define thermodynamic quantities of interest in the quantum realm, especially the entropy, an appropriate setup is needed to study how these definitions work within the nature of the quantum processes. In this regard, there are interesting works that try to shed light on the subject, such as the violation of the Clausius inequality for a quantum harmonic oscillator linearly coupled to a bath of oscillatory fields \cite{All2,Hil,Soltan}, studying the distinction between the Gibbs-preserving maps and thermal operations \cite{Fai}, Thermodynamic relations between two spatially separated entangled particles \cite{Soltan2} and the comparison of thermodynamic entropy of a quantum Brownian oscillator obtained by the partition function of the system with von Neumann entropy of the system \cite{Hor}.

Nowadays, it seems that the identity of the information is inseparable from its physical character. Quantum information has an important role in our understanding of quantum thermodynamics. Many thoughts emerge here to understand the fundamental relations between information and statistical mechanics. Accordingly, searching for thermodynamic rules in the quantum regime is vastly quantum information-oriented \cite{Gool}. This includes entropy measures for quantifying the uncertainties about events \cite{Cov}, emerging resource theories \cite{Fritz}, equilibration and maximum entropy principle \cite{Lind,Short,Cherian}, thermalization \cite{Riera} and entanglement theory in thermodynamics \cite{Horo}. Moreover, information theory suggests the emergence of Gibbs states from the quantum equilibration process within three conditions: equal a priory probability postulate, the assumption of weak coupling, and an assumption about the density of states of the bath that it grows faster than exponentially with energy and it can be approximated locally with an exponential \cite{Gool}. However, we discuss that the entropy maximization during the equilibration does not necessitate the emergence of Gibbs states.

Here, we study the thermodynamic behavior of a harmonic oscillator interacting with a thermal bath through a quantum circuit. The idea is to study how the thermal bath affects the internal energy and the entropy and to see how it can change the system by measuring it in the momentum space. To do this, we used the approach introduced by Binder and coworkers to formulate operational thermodynamics suitable for applying to an open quantum system undergoing a general quantum process, which can be described as a completely positive and trace-preserving (CPTP) map \cite{Bin}.

The paper is organized as follows. In section II, we review the derivation of the master equation for the model of quantum Brownian motion. We considered the conditions in which the Lindblad master equation holds. In section III, we briefly explain the operational first law of quantum thermodynamics and also discuss the requirements for a quantum equilibration to be considered as a thermalization process. In section IV, we solve the Lindblad master equation for a harmonic oscillator, which interacts with the thermal bath in a quantum circuit and is measured in the momentum space. Changes in the momentum probability pattern are illustrated and the entropy variation is calculated. In section V, we discuss how a quantum system that is going through an equilibration process does not necessarily end up in a thermal state. Also, by describing the role of entropy in the process, we discuss the concept of decoherence theory from a quantum thermodynamics point of view. Finally, in Section VI we show that in a system and thermal bath interaction a quantum equilibration cannot be considered as a thermalization process.

\section{Lindblad Model of Quantum Brownian Motion}

In this section, we review the Lindblad model of the quantum Brownian motion with the approach introduced by Maniscalco and coworkers \cite{Man}. The Master equation first was introduced by Gorini, Kossakowski, and Sudarshan in which the Born-Markov approximation leads to a master equation in a Lindblad form known as GKSL (Gorini-Kossakowski-Sudarshan-Lindblad) equation \cite{Gori}. In this system-reservoir model, the total Hamiltonian is defined by three parts
\begin{equation}
\label{totalH}
H=H_s+H_\varepsilon+H_{int},
\end{equation}
where $H_s$,$H_\varepsilon$, and $H_{int}$ are the Hamiltonians of the system, the environment, and the system-environment interaction, respectively. The central system is a quantum harmonic oscillator and the environment is a collection of harmonic oscillators as a thermal bath. Therefore, the total Hamiltonian can be written as
\begin{equation}
\label{TH}
H=(P^2+\Omega^2X^2)/2+\sum_i(p_i^2+\omega_i^2q_i^2)/2+H_{int},
\end{equation}
where $\Omega$ ($\omega_i$) is the system (the environment) frequency and $P$ and $X$ ($p_i$ and $q_i$) are momentum and position operators of the system (the environment), respectively. For simplicity, we write off the mass and consider $\hbar=1$. The form of the interaction between the system and the environment is such that the position coordinate $X$ of the central particle couples linearly to the positions $q_i$ of the thermal bath oscillators with coupling strength $c_i$. Thus the interaction Hamiltonian $H_{int}$ reads
\begin{equation}
\label{HInt}
H_{int}=X\otimes E=X\otimes\sum_i c_iq_i.
\end{equation}

Denoting $\rho$ as the total system-environment density matrix, the following assumptions are in order.  First, the system and the environment are supposed to be uncorrelated at t=0 which means \mbox{$\rho(0)=\rho_s(0)\otimes\rho_\varepsilon(0)$} with $\rho_s$ and $\rho_\varepsilon$ are the system and the environment density matrices, respectively. Second, we assume that the environment is stationary, that is $[H_\varepsilon,\rho_\varepsilon(0)]=0$ and also the expectation value of $E$ is zero, $\text{Tr}_E [E\rho_E(0)] = 0$. Finally, the system-environment coupling is weak and under the weak coupling, the factor of the oscillator frequency renormalization is negligible. Then, by averaging over the rapidly oscillating terms, one gets the following secular approximated master equation \cite{Man,Bre}
\begin{align}
\label{MaEq}
\frac{d\rho_s}{dt}=&-\frac{\Delta(t)+\gamma(t)}{2}[a^\dagger a\rho_s-2a\rho_s a^\dagger+\rho_s a^\dagger a] \nonumber \\
&-\frac{\Delta(t)-\gamma(t)}{2}[aa^\dagger \rho_s-2a^\dagger\rho_s a+\rho_s a a^\dagger],
\end{align}
where $a=(X+iP)/\sqrt{2}$ and $a^\dagger=(X-iP)/\sqrt{2}$ are the bosonic annihilation and creation operators, respectively. Also, the time dependent coefficient $\gamma(t)$ is responsible for classical damping and $\Delta(t)$ is a diffusive term. These coefficients are defined as
\begin{align}
\label{D}
\Delta(t)&=\int_0^t\kappa(\tau)\cos(\Omega\tau)\text{d}\tau, \\
\label{g}
\gamma(t)&=\int_0^t\mu(\tau)\sin(\Omega\tau)\text{d}\tau,
\end{align}
where
\begin{align}
\label{noise}
\kappa(\tau)=\sum_ic_i^2\langle \{q_i(\tau),q_i\}\rangle,
\end{align}
and
\begin{equation}
\label{dissipation}
\mu(\tau)=i\sum_ic_i^2\langle [q_i(\tau),q_i]\rangle,
\end{equation}
are noise and dissipation kernels, respectively.

The master equation (\ref{MaEq}) is similar to the Lindblad form but the coefficients are time-dependent. With the positive coefficients $\Delta(t)\pm\gamma(t)$ at all times, equation \eqref{MaEq} is a Lindblad-type master equation \cite{Man2}. Let us consider the case of an Ohmic spectral density for the reservoir with Lorentz-Drude cutoff \cite{Schl}
\begin{equation}
\label{sd}
J(\omega)=\frac{2\gamma_0\omega}{\pi}\frac{\Lambda^2}{\Lambda^2+\omega^2},
\end{equation}
where $\Lambda$ is the cut-off frequency and the dimensionless factor $\gamma_0$ describes the effective coupling strength between the system and the environment. Then, the coefficients $\Delta(t)$ and $\gamma(t)$ at the asymptotic long-time limit approach their stationary values, which their expression up to the second order in the coupling constant read as \cite{Man}
\begin{align}
\label{Delta}
\Delta &=\gamma_0^2\Omega\frac{r^2}{1+r^2}\coth(\frac{\Omega}{2kT}) \\
\label{Gamma}
\gamma &=\gamma_0^2\Omega\frac{r^2}{1+r^2},
\end{align}
where $k$ is Boltzmann constant, $T$ denotes temperature and $r=\Lambda/\Omega$. The master equation (\ref{MaEq}) becomes the well-known Markovian master equation of damped harmonic oscillator
\begin{align}
\label{ME}
\frac{d\rho_s}{dt}=&-\Gamma(\bar{n}+1)[a^\dagger a\rho_s-2a\rho_s a^\dagger+\rho_s a^\dagger a] \nonumber \\
&-\Gamma\bar{n}[aa^\dagger \rho_s-2a^\dagger\rho_s a+\rho_s a a^\dagger],
\end{align}
where $\Gamma=\gamma_0^2\Omega r^2/(1+r^2)$ and $\bar{n}=(e^{\Omega/kT}-1)^{-1}$. In this situation which the  coefficients $\Delta\pm\gamma$ are positive, the aforementioned relation \eqref{MaEq} is a Lindblad-type equation.


\section{Thermalization and Quantum Thermodynamic Equilibration}

\subsection{Thermalization Process \label{THreq}}

The idea of a thermodynamic system always approaching an equilibrium state in which it describes as if it occupied an immense multitude of microstates at the same time independent of the initial state is still puzzling \cite{Gem}. Thermalization is a straightforward process, a system in contact with a large enough thermal bath will reach an equilibrium state at the same temperature. Furthermore, the thermalization process contains many aspects that we can decompose into the following postulates \cite{Gem,Lind}:

1.~{\it Equilibration}: A system equilibrates if it extends towards some particular state and remains in that state for all times. Equilibration does not concern what the equilibrium state is and whether it depends on the initial state or not as long as it is stationary.

2.~{\it Bath state independence}: Since the equilibrium state of the system is controlled by the contact with a heat bath, the only specifying parameter is its temperature. When the system reaches equilibrium, the equilibrium state should depend only on the temperature of the bath.

3.~{\it Subsystem state independence}: If the subsystem is small in comparison to the bath, the equilibrium state of the subsystem should be independent of its initial state.

4.~{\it Boltzmann form of the equilibrium state}: With certain conditions on the interaction Hamiltonian, the equilibrium state of the subsystem can be written in the Boltzmannian form $\rho_s=\frac{1}{Z}\exp(-\frac{H_s}{k_BT})$.

In the following, we need to understand the various aspects of a quantum thermodynamic process to argue, {\it can we comprehend quantum equilibration as a thermalization process in its nature?}

\subsection{Operational First Law of Quantum Thermodynamics}

The first law of thermodynamics discusses that the changes of the internal energy $(\Delta E)$ consist of two terms: work ($W$) and heat ($Q$). To explain the aspects of the first law in quantum thermodynamics we consider a cyclic process. Here, we call a process cyclic, in which the Hamiltonian of the system at the beginning and the end of the process are identical \cite{Bin}.

In a system with the Hamiltonian
\begin{align}
\label{HT}
H=\sum \varepsilon _n \vert \varepsilon _n \rangle \langle \varepsilon _n\vert \hspace{0.4cm} \text{;} \hspace{0.4cm} \varepsilon _{n+1}\geq \varepsilon _n \forall n,
\end{align}
the internal energy defines as $E(t)=\tr [\rho (t) H(t)]$. For the density matrix we have
\begin{align}
\label{DT}
\rho=\sum r_n \vert r_n\rangle\langle r_n\vert \hspace{0.5cm} \text{;} \hspace{0.4cm} r_{n+1}\leq r_n \forall n.
\end{align}

By extracting the maximum work via a cyclic unitary process from a system, the system density matrix of the system in Eq. \eqref{DT} ends up in a passive state $\pi$. passive states are diagonal in the basis of the Hamiltonian \eqref{HT} with decreasing populations for increasing energy levels, expressed as
\begin{align}
\label{PT}
\pi=\sum r_n \vert \varepsilon_n\rangle\langle \varepsilon_n\vert \hspace{0.5cm} \text{;} \hspace{0.4cm} r_{n+1}\leq r_n \forall n.
\end{align}
Passive states are in a complete agreement with the second law of thermodynamics in Kelvin-Planck formulation. It is of a note that Gibbs states (thermal states) are consequently passive states \cite{Per}. The process of unitary cyclic work extraction from a non-passive state \eqref{DT}, with respect to the Hamiltonian \eqref{HT} is called ergotropy \cite{Bin,Pus}
\begin{align}
\mathcal{W}=\tr\big[\rho H-\pi H\big]=\sum_{m,n} r_m \varepsilon_n \big[\vert  \langle \varepsilon_n \vert r_m\rangle\vert^2- \delta_{mn}\big]
\end{align}

Now we consider a non-cyclic unitary process in which the initial and final Hamiltonians $H$ and $H'$ are different where $H'=\sum \varepsilon' _n \vert \varepsilon' _n \rangle \langle \varepsilon' _n\vert$ with $\varepsilon' _{n+1}\geq \varepsilon' _n$. Also we consider the process adiabatic. The initial state $\pi_m$ and the final state $\pi'$ are passive states respect to $H$ and $H'$. Since the evolution is unitary, there is no heat transfer and any change in the internal energy is due to the adiabatic work which is defined as
\begin{align}
\label{Wad}
\langle W\rangle_{\ad}=\tr\big[\pi'H' \big]-\tr \big[\pi_mH\big].
\end{align}
In a general quantum process $(\rho,H)\longmapsto (\rho',H')$, the changes of internal energy is given by
\begin{align}
\label{intE}
\Delta E=\tr\big[\rho' H'\big]-\tr\big[\rho H\big].
\end{align}
Regarding the concepts of ergotropy and adiabatic work and considering $\pi_m=\sum r'_n \vert \varepsilon_n\rangle\langle\varepsilon_n\vert$, the operational first law of quantum thermodynamics introduced as \cite{Bin}
\begin{align}
\label{FL}
\Delta E= \Delta \mathcal{W}+\langle W\rangle_{\ad}+\langle Q\rangle_{\op},
\end{align}
where $\Delta\mathcal{W}=\mathcal{W}(\rho',H')-\mathcal{W}(\rho,H)$ and $\langle Q\rangle_{\op}=\tr\big[\pi_m H\big]-\tr\big[\pi H\big]$ known as operational heat. Hence, any quantum thermodynamic process that can be described with a CPTP (complete positive and trace-preserving) map, obeys operational first law of quantum thermodynamics \eqref{FL}.

\subsection{Quantum Equilibration vs Quantum Thermalization \label{QET}}

In a thermodynamic process that a quantum system interacts with a thermal bath, the system reaches an equilibrium state usually with no change in system Hamiltonian. This process is quantum equilibration which is commonly known as quantum thermalization. Here, we discuss that can we consider quantum equilibration and quantum thermalization generally identical in such cases or not.

In the process of system interaction with a thermal bath, the system Hamiltonian remains unchanged. Regarding equation \eqref{Wad} the adiabatic work is zero ($\langle W\rangle_{\ad}=0$). In a thermalization process the system always ends up in a thermal state which is a passive state, thus no work can be extracted via a cyclic unitary process from a thermal state. Accordingly, the operational first law of quantum thermodynamics \eqref{FL} for any thermalization process reduces to
\begin{align}
\label{TFL}
\Delta E_{th}=\mathcal{W}(\rho,H)+\langle Q\rangle_{\op},
\end{align}
since $\mathcal{W}(\rho',H)$ is zero as the final state $\rho'$ itself is a passive state. On the other hand in an equilibration process as a general process without any changes in Hamiltonian we have
\begin{align}
\label{EFL}
\Delta E_{eq}=\Delta \mathcal{W}+\langle Q\rangle_{\op},
\end{align}
with $\Delta\mathcal{W}=\mathcal{W}(\rho',H)-\mathcal{W}(\rho,H)$. This shows us that we can call an equilibration process, a thermalization, in which we can extract no work from the equilibrium state via a cyclic unitary process ($\mathcal{W}(\rho',H)=0$).


\section{System-Bath Interaction in a Quantum Circuit}
\begin{figure}
\centering
\begin{quantikz}
\lstick{\ket{0}} & \gate[style={fill=yellow!50}]{H} & \gate[style={fill=yellow!50}]{\phi} & \gate[style={fill=red!35}]{\begin{array}{c}
\text{Thermal Bath} \\ (\rho_{th})
\end{array}} & \gate[style={fill=yellow!50}]{H} & \meter[style={fill=brown!50}]{\footnotesize Pr$(P)$ \\ \footnotesize Pr$(X)$}
\end{quantikz}
\caption{A particle in the state $\ket{0}$ enters the Hadamard gate. then, after passing through the phase shift gate, it interacts with the thermal bath. Afterward, the particle enters to the second Hadamard gate and reaches the detector that measures in the momentum/position space.}\label{case01}
\end{figure}
In this section, we study a two-state ($\vert 0\rangle$ and/or $\vert 1\rangle$) harmonic oscillator interacting with a collection of harmonic oscillators as a thermal bath in the quantum circuit represented in FIG. \ref{case01}. The harmonic oscillator in the state of $\ket{0}$ passes through a Hadamard and a phase gates, ends up in the state $\frac{1}{\sqrt{2}}(\vert 0\rangle+e^{i\phi}\vert 1\rangle)$ before it reaches the thermal bath. Therefore, for the density matrix of the system just before the interaction with the thermal bath we have
\begin{equation}
\label{c1-idm}
\begin{pmatrix}
1/2 & e^{-i\phi}/2 \\
e^{i\phi}/2 & 1/2
\end{pmatrix}.
\end{equation}
The dynamics of the decoherence effect is determined by the master equation \eqref{ME}, that can be solved under the rotating wave approximation, wherein we assume that the contribution of terms $a^2$ and $a^{\dagger2}$ is negligible. So the annihilation operator is defined as $a\approx\vert 0\rangle\langle 1\vert$. Here, the central system is represented by a two-state harmonic oscillator using the rotating-wave approximation, which is mathematically identical with a qubit (spin 1/2 system) \cite{Bre,Schl}. One can show that the state of the system after the interaction with the thermal bath is described by the following density matrix
\begin{equation}
\label{c1-tdm}
\begin{pmatrix}
\frac{2(\bar{n}+1)-\eta^2}{2(2\bar{n}+1)} & \frac{1}{2}\eta e^{-i\phi} \\
\frac{1}{2}\eta e^{i\phi} & \frac{2\bar{n}+\eta^2}{2(2\bar{n}+1)}
\end{pmatrix},
\end{equation}
where $\eta=e^{-\Gamma t(2\bar{n}+1)}$. Accordingly, the thermal bath shows its dissipative effects on the system. It might seem that the thermal bath leaves the system in a mixed state at first glance, but we will discuss in the next section that the state \eqref{c1-tdm} could be entirely coherent in certain conditions. After passing the system through the second Hadamard gate, the final state of the system is
\begin{equation}
\label{c1-fdm}
\rho_s(t)=
\begin{pmatrix}
\frac{1+\eta\cos\phi}{2} & (\frac{1-\eta^2}{2(2\bar{n}+1)})+\frac{\eta\sin\phi}{2} \\
(\frac{1-\eta^2}{2(2\bar{n}+1)})+\frac{\eta\sin\phi}{2} & \frac{1-\eta\cos\phi}{2} 
\end{pmatrix}.
\end{equation}

 The master equation \eqref{ME} is well-known and broadly used for qubit dynamics up to this day \cite{Bre,Vacc,Hall}. Now we used this master equation for system-bath interaction in a quantum circuit that acts as an interferometer. After the interaction, the non-diagonal elements of the density matrix approach to zero ($\eta\rightarrow 0$ in equation \eqref{c1-tdm}). But this does not necessarily show the coherence is lost. The second Hadamard in the circuit creates the interference terms, following us to observe the system's coherence through the changes in the interference patterns (non-diagonal elements are not zero at $\eta\rightarrow 0$ in equation \eqref{c1-fdm}). Also, we are going to track the thermodynamic parameters (especially entropy) through this process and search for the roots of coherence loss in quantum thermodynamics.

\begin{figure}
\centering
\includegraphics[scale=0.45]{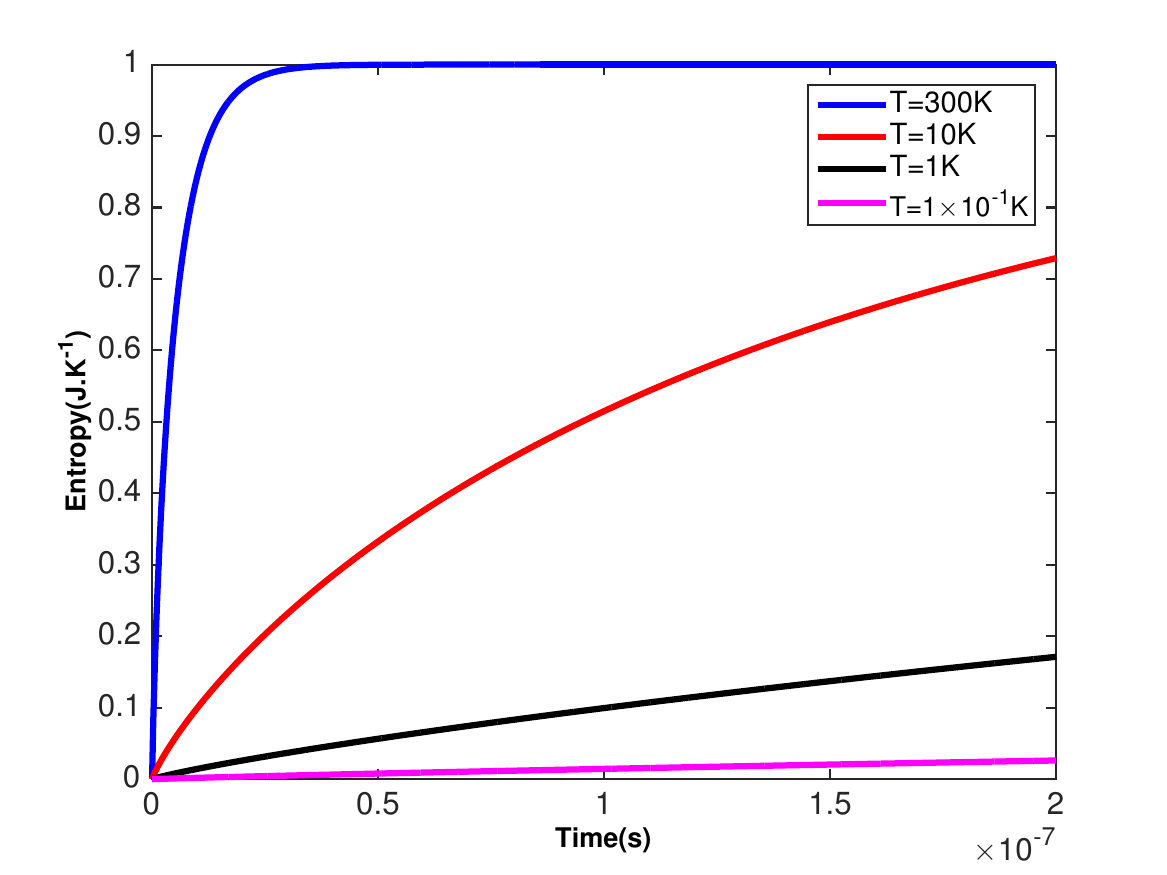}
\caption{Entropy changes of the system versus time in a variety of temperatures. $\Omega=1\times10^{12}s^{-1}$} \label{DS}
\end{figure}

The von Neumann entropy $S$ of the system is given by
\begin{align}
\label{entropy}
S(t)=-&\sum_{i=0,1}[(\frac{1}{2}+(-1)^i\frac{1}{2}\sqrt{\frac{1-\eta^2(2-\eta^2-(2\bar{n}+1)^2)}{(2\bar{n}+1)^2}}) \nonumber \\
&\times\log(\frac{1}{2}+(-1)^i\frac{1}{2}\sqrt{\frac{1-\eta^2(2-\eta^2-(2\bar{n}+1)^2)}{(2\bar{n}+1)^2}})],
\end{align}
where $S(0)=0$, so we have $\Delta S(t)=S(t)$. As we expect from (\ref{Slaw}), FIG. \ref{DS} shows that as time goes on, in room temperature, entropy rapidly increases till the decoherence begins to work ($\eta\rightarrow 0$). Then, system reaches the equilibrium state and entroby tends a constant value (nearly equal to $1$ but not exactly), where we expect to see the quantum effects are disappeared. The process happens slower in lower temperatures as is apparent in FIG. \ref{DS}.

After the particle passed the second Hadamard gate we measure the system in the momentum/position space. The probability distribution of diagonal and off-diagonal elements of $\rho_s(t)$ in  (\ref{c1-fdm}) can be obtained as $\text{Pr}(X)=\langle X\vert\rho_s(t)\vert X\rangle$ and $\text{Pr}(P)=\langle P\vert\rho_s(t)\vert P\rangle$, respectively. By using the well-known harmonic oscillator wave functions as $\langle X\vert 0\rangle=\Psi_0(X-d/2)$ and $\langle X\vert 1\rangle=\Psi_1(X+d/2)$ and also their Fourier transform in the momentum space we simply can calculate the momentum distribution (interference fringes). Here, $d$ represents the optical path length. We expect the well-separated two peaks along $X$ axis, representing the pointer positions of the detector for $\ket{0}$ and $\ket{1}$ states. For the system described by the density matrix (\ref{c1-fdm}), the probability distribution along the $P$ axis can be depicted by the following relation
\begin{align}
\label{ip}
\text{Pr}(P)=\sqrt{\frac{1}{\Omega\pi}}e^{-\frac{P^2}{\Omega}}[1&+\frac{\eta^2-1}{2\bar{n}+1}\sin(Pd) \nonumber \\
&+\eta\sin\phi\cos(Pd)],
\end{align}

\begin{figure}
\centering
\includegraphics[scale=0.57]{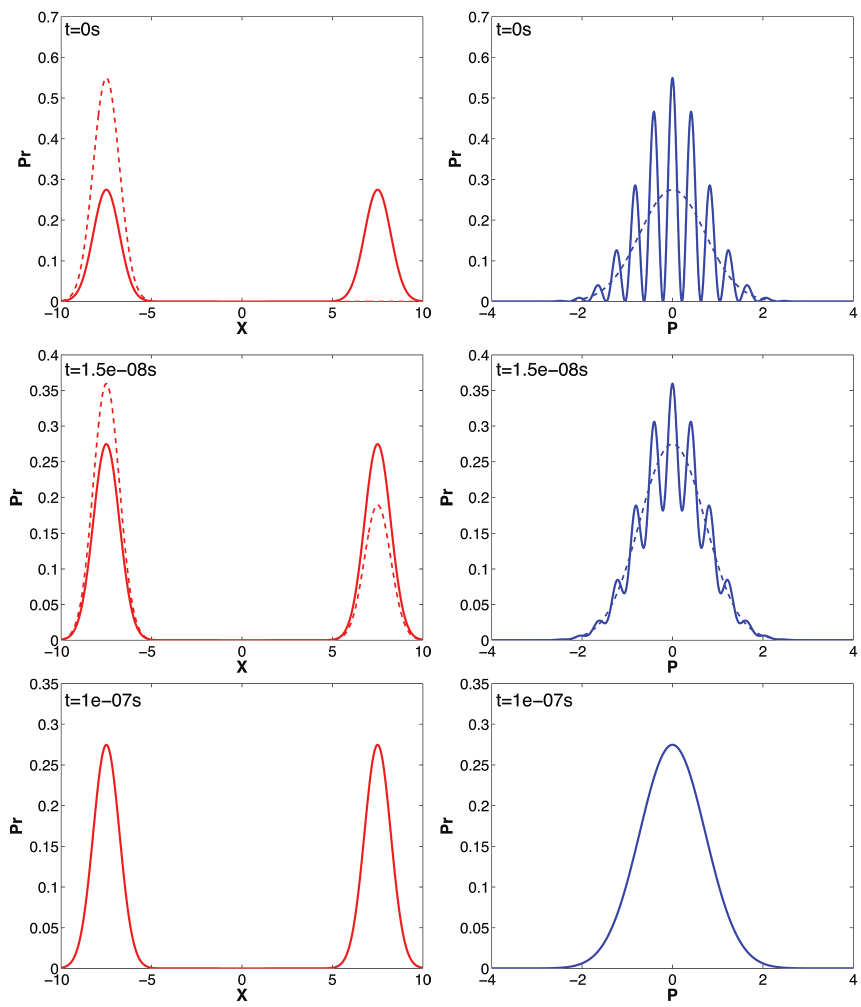}
\caption{The position (left side) and the momentum (right side) distribution functions of the system at $t=0s$, $1.5\times10^{-8}s$ and $1\times10^{-7}s $ with $\Omega/T=10^{10}s^{-1}K^{-1}$. The solid line shows the evolution for the phase factor $\phi=\pi/2$, and the dashed line is for $\phi=0$.}\label{IE}
\end{figure}

Two important terms appear in the momentum probability (\ref{ip}). First the term including "$\eta\sin\phi\cos(Pd)$" which has the main role in the appearance of the interference pattern. By passing time, $\eta$ tends to zero and as we expect, the interference pattern vanishes after a certain time known as decoherence time, as is shown in FIG. \ref{IE}. Moreover, choosing the phase factor $\phi=0$, the mentioned term is zero again and no interference appears. However, as we see in FIG. \ref{IE}, at the beginning time ($t=0$) with $\phi=0$, the pointer position is fixed and no interference pattern is observed before the decoherence works.

The second term "$\frac{\eta^2-1}{2\bar{n}+1}\sin(Pd)$" is at $t=0$ (with $\eta\approx1$) negligible compared to the term mentioned above because $\bar{n}$ is in the denominator. This term has also a definite contribution to the appearance of the interference pattern, though it is partial. Yet, when time passes and the decoherence begins to work and/or one chooses the phase factor $\phi$ to be zero, its contribution could be significant, especially in situations where $\bar{n}$ is small.

\section{Results and Discussion}

Let us discuss and conclude the fundamentally important effects of the thermal bath on the system. If one considers the interference pattern equation (\ref{ip}) for $\phi=0$ or after the decoherence process when $\eta=0$, in both of these situations the term "$\frac{\eta^2-1}{2\bar{n}+1}\sin(Pd)$" will have a significant role in the appearance of the interference pattern, especially for small values of $\bar{n}$. The condition $\phi=0$ helps us to compare the problem with the case in which we have a closed system with no interference. Also, $\eta=0$ makes us sure that we have the most influence of the thermal bath on the system. With $\phi=0$, we just have the term "$\frac{\eta^2-1}{2\bar{n}+1}\sin(Pd)$", which is zero at $t=0$, but when time goes on, especially after the decoherence process, it reaches the maximum value of "$\frac{-1}{2\bar{n}+1}\sin(Pd)$".

Let us now consider the entropy function in such a situation. After the decoherence process, entropy tends to the following constant value
\begin{equation}
\label{emax}
S(t\rightarrow\infty)=-\frac{\bar{n}+1}{2\bar{n}+1}\log_2\frac{\bar{n}+1}{2\bar{n}+1}-\frac{\bar{n}}{2\bar{n}+1}\log_2\frac{\bar{n}}{2\bar{n}+1}.
\end{equation}
For large values of $\bar{n}$ (at high temperatures, or more precisely, low $\Omega/T$), the entropy value approaches $1$. When a system is in a thermal state with equal probabilities of its eigenvectors (completely mixed state), the entropy has a maximum value, which for the two-dimensional Hilbert state is $1$. FIG. \ref{IE} shows us that the decoherence makes this possible that the system has a chance to be found in another state (i.e., $\vert0\rangle$) too since both detectors have the same chance to click when the decoherence process is completed. Yet, when $\Omega/T$ is large enough, $S(t\rightarrow\infty)$ is far from $1$ and the state of the system is not completely mixed, as is described by the interference pattern equation (\ref{ip}). Even after the decoherence process is completed, high values of $\Omega/T$ cause the interference pattern to appear, due to the term "$\frac{-1}{2\bar{n}+1}\sin(Pd)$". This is interesting because we have the perception that an equilibrium state can be described by a thermal state with classical behavior. Traditionally, during decoherence, we expect that the state goes into a thermalization process and ended in a mixed thermalized state as an equilibrium state and loses its coherence (interprets as a quantum to classical transition). Thus, we do not expect to see any coherence in the system. However, the observation of interference fringes after the decoherence process reveals that the system is coherent still, which is in contrast with the classical interpretation of the equilibrium state. \cite{Jeo,Jeo2}. This indeed changes our attitude about the thermalized state as a quantum state in equilibrium. Somehow, this shows that the system affected by the environment keeps some of its quantum properties. So, it seems that actually, the maximization of the entropy leads the system to a new equilibrium state that cannot always be described by Gibbs states.

\begin{figure}
\centering
\includegraphics[scale=0.45]{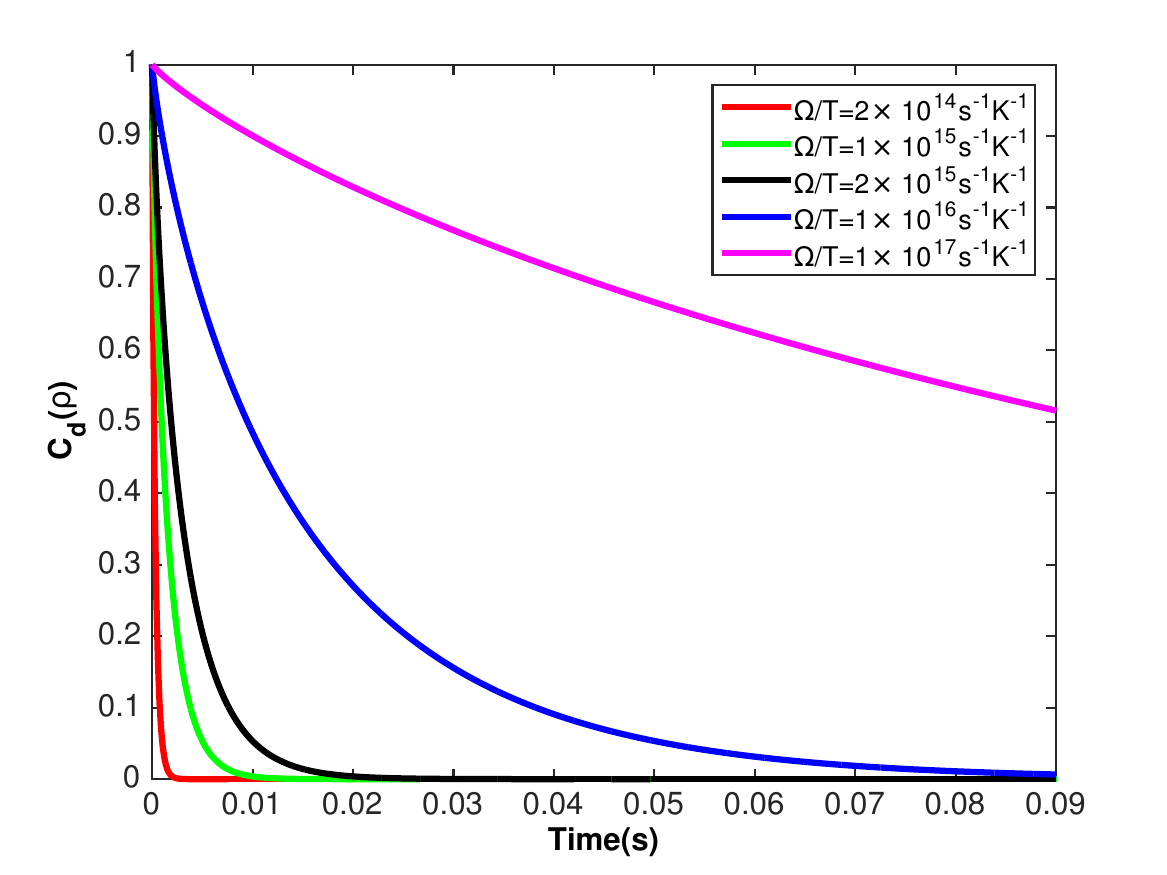}
\caption{The distillable coherence $C_d(\rho)$ versus time with different values of $\Omega/T$. As time goes on, the larger $\Omega/T$ causes the distillable coherence decreases with a lower slope.} \label{Cdf}
\end{figure}

\begin{figure}
\centering
\includegraphics[scale=0.36]{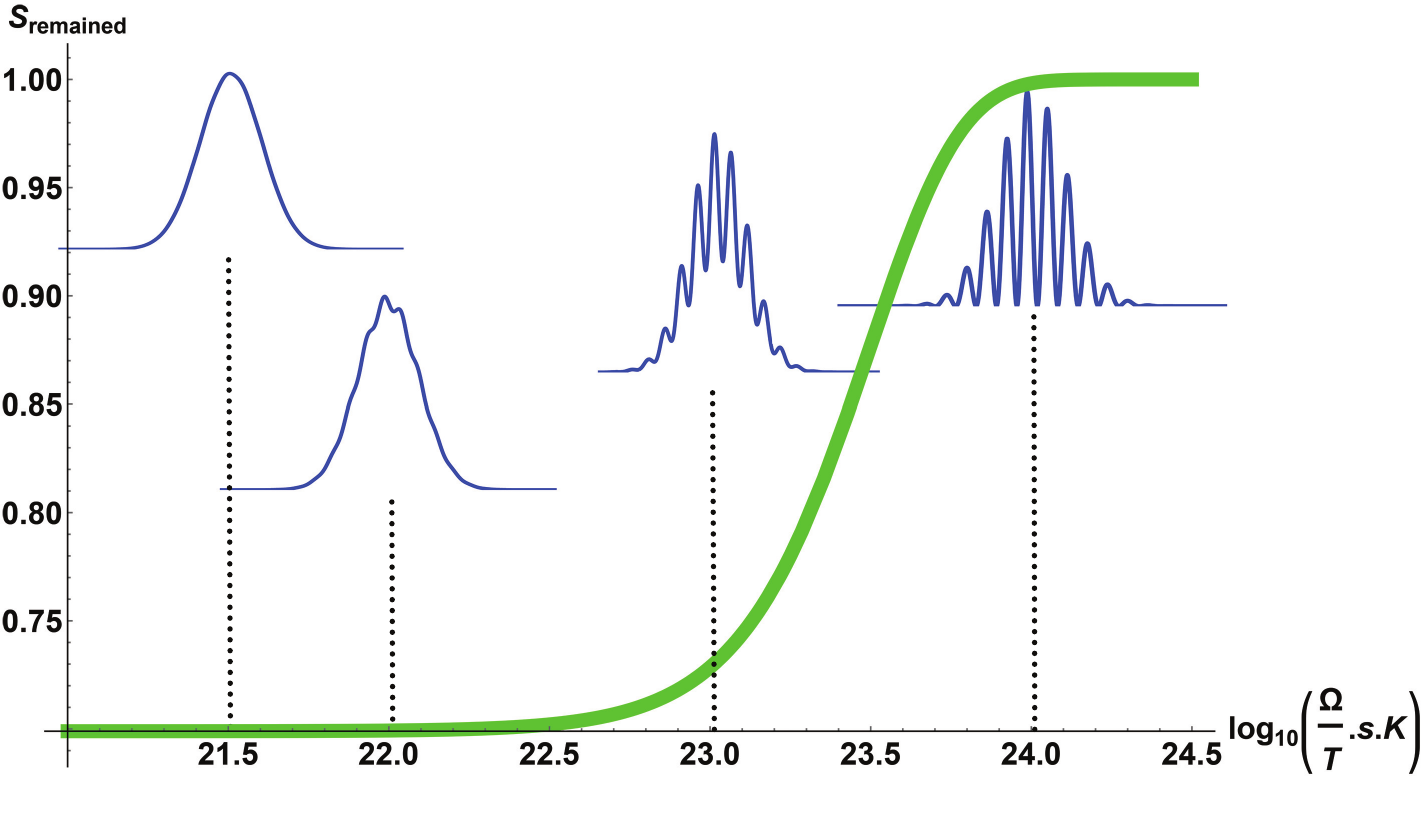}
\caption{Remained entropy ($S_{rem}$) changes versus $\log(\frac{\Omega}{T}.s.K)$. The interference patterns are illustrated among the remained entropy with $\log(\frac{\Omega}{T}.s.K)=21.5$, $22$, $23$ and $24$. The phase factor $\phi$ is taken zero here.} \label{Er}
\end{figure}

For investigating the system coherence in this setup, one can use the distillable coherence $C_d(\varrho)$ as a quantifier \cite{Win}. The distillable coherence is the optimum number of maximally coherent states which can be obtained from a state $\varrho$ using incoherent operations. It has been shown that the distillable coherence can simply write as \cite{Win,Str}
\begin{equation}
\label{Cdeq}
C_d(\varrho)=S(\Xi[\varrho])-S(\varrho),
\end{equation}
where $\Xi[\varrho]=\sum_{i=0}^{d-1}\vert i\rangle\langle i\vert\varrho\vert i\rangle\langle i\vert$ is the dephasing operator. In the case of the density matrix, $\rho$ mentioned in equation \eqref{c1-fdm}, for the entropy of the dephased density matrix we have
\begin{align}
\label{dephased}
S(\Xi[\rho])=-&\sum_{i=0,1}[\frac{1+(-1)^i\eta\cos\phi}{2} \nonumber \\ &\times\log(\frac{1+(-1)^i\eta\cos\phi}{2})].
\end{align}
Therefore, because of \eqref{entropy}, the distillable coherence $C_d(\rho)$ can be calculated for the system with different $\Omega/T$. As we expected, the value of the $C_d(\rho)$ decreases as time goes on. However, as FIG. \ref{Cdf} shows the large values of $\Omega/T$ cause the system remains coherent for a long time. If one considers $\Omega/T$ to be large enough, the distillable coherence almost remains invariant and the system stays coherent. So, we can observe the interference patterns after the decoherence process, as is represented in FIG. \ref{Er}.  Since the system remains coherent, the concept of the decoherence theory becomes ambiguous. Also, the role of thermodynamic parameters makes us search for different causes for such behavior. In this regard, we study the classicality of the system's final state among the changes in thermodynamic parameters.

\begin{figure}
\centering
\subfigure[]{\label{mixednessfig}
\includegraphics[scale=0.35]{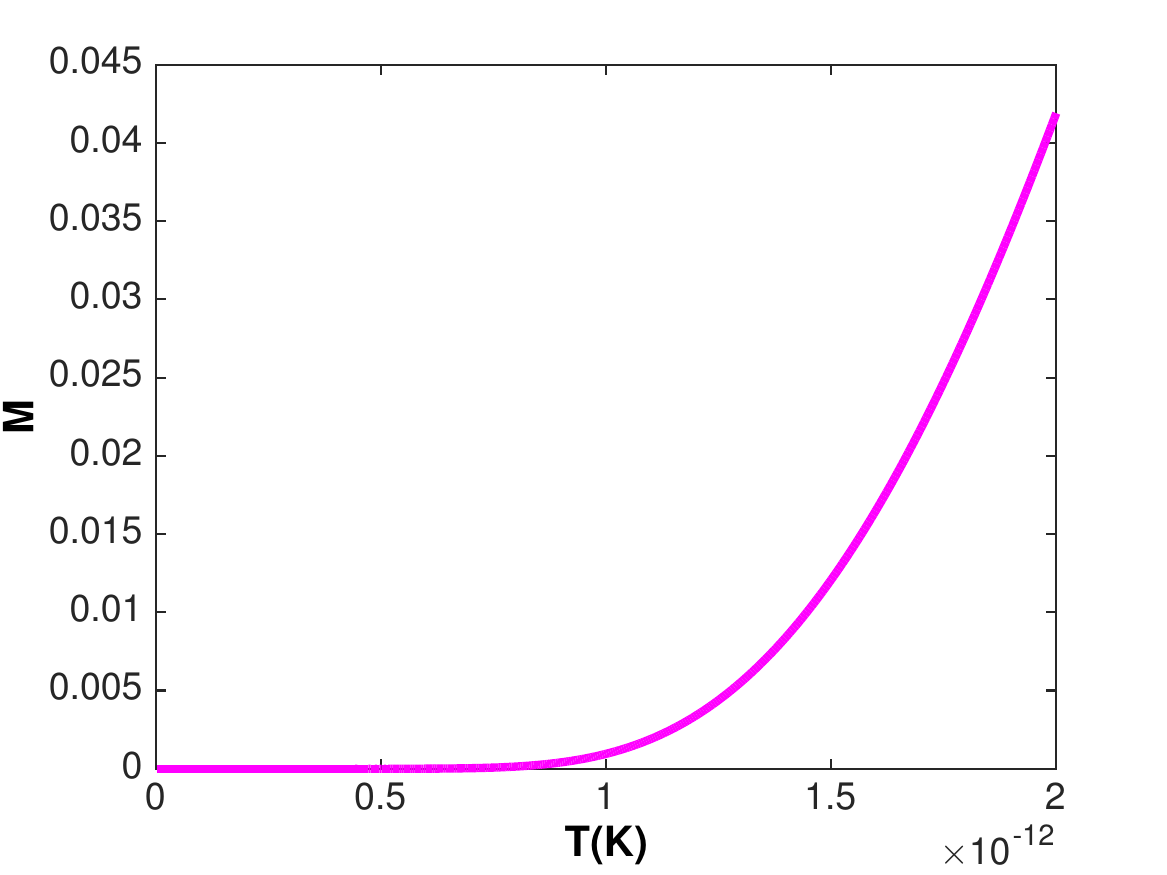}}
\subfigure[]{\label{visibilityfig}
\includegraphics[scale=0.35]{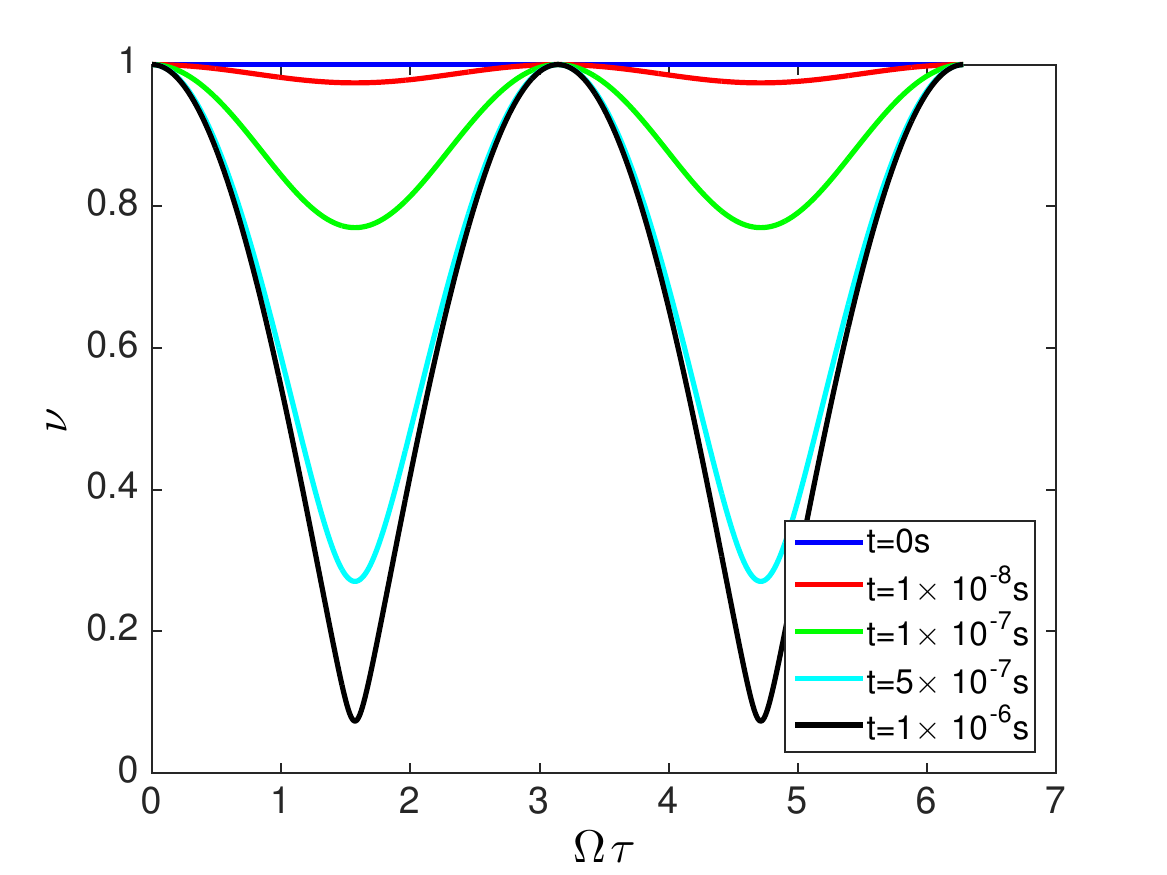}}
\caption{(a) The changes of the degree of mixedness ($M$) versus temperature. The system remains coherent at low temperatures (b) The interference fringes visibility versus the time interval ($\tau$) as the decoherence process evolves with time. We considered $\Omega=1\times 10^{12}s^{-1}$.}
\label{classicality}
\end{figure}

To examine the classicality of the resulted state we calculate the degree of mixedness ($M=1-\text{Tr}[\rho^2]$) of the system exactly after its interaction with thermal bath with the density matrix \eqref{c1-tdm} in the limit of $t\rightarrow\infty$ or $\eta\rightarrow 0$. Fortunately, the degree of the mixedness of the system is near zero at low temperatures which shows that the system is completely coherent, although its density matrix is diagonal. FIG. \ref{mixednessfig} shows the changes of $M$ versus temperature in which with the increasing of the temperature the system ends up in the thermal state as the degree of mixedness approaches $M=0.5$. Furthermore, the visibility $v$ of the interference fringes (FIG. \ref{Er}) is defined as \cite{Walls} $v=(I_{max}-I_{min})/(I_{max}+I_{min})$ where $I_{max (min)}$ is the maximum (minimum) intensities on the detector. We sketched $v(\tau)$ along with time in FIG. \ref{visibilityfig}. $\tau$ represents the time interval between the measurements. As we expect, the visibility is $1$ before the decoherence process starts. Interestingly, as FIG. \ref{visibilityfig} shows, by setting an appropriate time interval we could reach the visibility near $1$ even when the decoherence process is completed. It is obvious that the interference fringes with high visibility are incompatible with classical physics and evidence of quantum coherence \cite{Jeo}.

Here, we introduce a measure to see how much the equilibrium state differs from a thermal state and keeps its quantum properties. This helps us to have a better understanding of the emergence of classical traits. This is based on the distance of the system's maximum entropy with the entropy of the system in a completely mixed state which we call the remained entropy, $S_{rem}$:
\begin{equation}
\label{Entre}
S_{rem}:=S_{cm}-S(t\rightarrow\infty),
\end{equation}
where $S_{cm}$ is the entropy of a completely mixed state (thermal state with equal probabilities for its eigenvectors) and its value is $1$ in our case. The remained entropy is the difference between two von Neumann entropies. As the entropy production is the main reason for the loss of coherence, the non-zero value of the remained entropy ensures us that the system is still coherent. The greater the amount of the remained entropy is the more quantum properties the system shows. In this regard, FIG. \ref{Er} shows that with growing the remained entropy, the interference pattern due to the effects of the environment (caused by the term "$\frac{-1}{2\bar{n}+1}\sin(Pd)$" ) appears more distinctly. The remained entropy shows the distance of the thermalized state from the classical mixture which does not show any coherence on any basis. However, by getting away from a classical mixture, we hope to see the coherence. As we expect, with higher remained entropy, the distillable coherence decays slower. The observations here bring us the idea of the thermodynamic control of open quantum systems. Designing systems with slow entropy productions can revolutionize quantum technologies. The remained entropy is a perfect measure to track the quantum properties of the systems in such situations. In the upcoming section, we accurately evaluate the requirements of an equilibration process to be considered a thermalization process. We show that in a discussed system-thermal bath interaction, thermalization has not occurred.
\\

\section{can we comprehend quantum equilibration as a thermalization process in its nature?}

In the previous section, we discussed that the final state of the system interacting with a thermal bath behaves differently from what we expected as a thermal state, especially in lower temperatures. Now we analyze the equilibration process of the system to see is it a thermalization in its nature or not.

As we discussed in section \ref{QET}, a thermal state is a passive state and we cannot extract work from it using a cyclic unitary evolution. The master equation (\ref{ME}) generates trace-preserving completely positive dynamics. Thus, the equilibrium state in a thermalization process must have zero ergotropy (according to \eqref{TFL} and \eqref{EFL}). Considering equation \eqref{c1-fdm} after the equilibration process ($\eta\rightarrow 0$) as the equilibrium state, by extracting maximum work with a cyclic unitary process, system ends up in the passive state below
\begin{align}
\pi_s=
\begin{pmatrix}
\frac{\bar{n}+1}{2\bar{n}+1} & 0 \\
0 & \frac{\bar{n}}{2\bar{n}+1} 
\end{pmatrix}.
\end{align}
Though, the extracted ergotropy is equal to
\begin{align}
\label{exW}
\mathcal{W}=\tr\big[\rho_sH_s-\pi_sH_s\big]=-\frac{\Omega}{2(2\bar{n}+1)}.
\end{align}
$\mathcal{W}$ only is zero in large values of $\bar{n}$ and thereupon in high temperatures or low system frequencies.. Since the value of the extracted ergotropy is not zero, we can confidently conclude that quantum thermodynamics equilibration is not a thermalization process in its nature. According to relations \eqref{TFL},\eqref{EFL} and \eqref{exW}, we have
\begin{align}
\Delta E_{th}\neq \Delta E_{eq}.
\end{align}
In other words, when a quantum system is coupled to a thermal bath, an equilibration process occurs and it is not necessarily a thermalization. Unless at high temperatures the extracted ergotropy \eqref{exW} tends to zero and the equilibrium state tends to a thermal state too.

The final equilibrium state is strongly dependent on the central system frequency ($\Omega$). Also, In high-frequency systems or low temperatures, the system state is not in the form of Boltzmann state as we discussed in the previous section. Accordingly, the discussed process conflicts with the third and fourth requirements of a thermalization process mentioned in section \ref{THreq}.

\subsection*{General Model}
Now, let's have a look on general model in an N-dimensional Hilbert space. For an N-level system, we can write master equation \eqref{ME} as
\begin{widetext}
\begin{align}
\label{NME}
\sum_{n=0}^{N-1}\sum_{m=0}^{N-1}\frac{\text{d}C_{nm}}{\text{d}t}\vert n \rangle\langle m \vert=&-\Gamma(\bar{n}+1)\big{[}\sum_{n=0}^N\sum_{m=0}^{N-1} C_{nm}n\vert n \rangle\langle m \vert-2\sum_{n=0}^{N-2}\sum_{m=0}^{N-2} C_{n+1m+1}\sqrt{(n+1)(m+1)}\vert n \rangle\langle m \vert \nonumber \\
&+\sum_{n=0}^{N-1}\sum_{m=0}^{N-1} C_{nm}m\vert n \rangle\langle m \vert \big{]}-\Gamma\bar{n}\big{[} \sum_{n=0}^{N-2}\sum_{m=0}^{N-1} C_{nm}(n+1)\vert n \rangle\langle m \vert \nonumber \\
&-2\sum_{n=0}^{N-1}\sum_{m=0}^{N-1} C_{n-1m-1}\sqrt{nm}\vert n \rangle\langle m \vert+\sum_{n=0}^{N-1}\sum_{m=0}^{N-2} C_{nm}(m+1)\vert n \rangle\langle m \vert \big{]},
\end{align}
\end{widetext}
where, $C_{nm}$s represent the elements of the density matrix $\rho_s(t)$. One can show that the answer to the equation \eqref{NME} consists of two parts. The first part that results from the terms containing $C_{nm}$, is in the form of $A_{nm}\exp[-\Gamma N(\bar{n}+1/2) t]$, where $A_{nm}$ is a constant. Terms containing $C_{n-1m-1}$ and $C_{n+1m+1}$ result in the second part that is in the form of $B_{nm}(\bar{n})$. The value of $B_{nm}$s depend on initial conditions as it is apparent from equation \eqref{NME}. Thus, the elements of the density matrix $\rho_s(t)$ are in the form of
\begin{align}
\label{Cnm}
C_{nm}(t)=B_{nm}(\bar{n})+A_{nm}(\bar{n})e^{-\Gamma N(\bar{n}+1/2) t}.
\end{align}

For diagonal elements, By applying $\text{Tr}[\rho_s(t)]=\sum_{n=0}^{N-1}C_{nn}=1$ in $t=0$ and $t\rightarrow\infty$ we have
\begin{align}
\label{trcon1}
&t=0 ~~ \Longrightarrow ~~ \sum_{n=0}^{N-1}B_{nn}=1 \\
\label{trcon2}
&t\rightarrow\infty ~~ \Longrightarrow ~~ \sum_{n=0}^{N-1}A_{nn}=0.
\end{align}
The coefficients $B_{nn}$ are of form
\begin{align}
\label{Bnn}
B_{nn}=\frac{b_{nn}}{N(\bar{n}+1/2)},
\end{align}
where $b_{nn}$s are coefficients with $\bar{n}$ dependency, staisfying the following relation
\begin{align}
\label{sumbnn}
\sum_nb_{nn}=N(\bar{n}+1/2).
\end{align}
On the other hand, non-diagonal elements decay to "zero" at long times. Thus, we expect for non-diagonal elements that $B_{mn}=0$. In this regard, we have $A_{nm}=C_{nm}(0)$ for $n\neq m$.

Now we study the mixedness ($M=1-\text{Tr}[\rho^2]$) for this general model in low- temperature and high-temperature conditions. According to \eqref{Cnm} and \eqref{Bnn}, after a long time ($t\rightarrow\infty$) we have
\begin{align}
\label{MGm}
M=1-\sum_n\frac{b_{nn}^2}{N^2 (\bar{n}+1/2)^2}.
\end{align}
Using Cauchy-Schwarz inequality, one can show that
\begin{align}
\label{C-S}
\sum_nb_{nn}^2\leqslant N^2 (\bar{n}+1/2)^2.
\end{align}
The relations \eqref{MGm} and \eqref{C-S} perfectly show us that in low temperatures ($\bar{n}$ is too small) the value of mixedness tends to "zero", where shows that the system remains coherent. On the other hand, at high temperatures where the value of $\bar{n}$ is high, we have $M\rightarrow 1$ that shows the state of the system is nearly mixed. The same scenario happens for remained entropy. At high temperatures $B_{nn}$s tend to $1/N$ and the remained entropy reaches zero, consequently. Decreasing the temperature increases the value of the remained entropy. 

\section{Conclusion}

The main objective of the current paper is to study the requirements of the quantum equilibration process to be considered a thermalization process. In this regard, we used the operational first law of quantum thermodynamics to explain the thermodynamic differences between quantum equilibration and thermalization processes. Also, This purpose leads us to investigate the variations of the system's coherence in search of the footprints of responsible thermodynamical parameters in the system's evolution.

We studied a two-state harmonic oscillator in a superposition state that interacts with a thermal bath of oscillatory fields in a quantum circuit. Later, the particle passed through a Hadamard gate, and by measuring it in the momentum space, we illustrated the interference pattern (momentum probability distribution) in a variety of temperatures. We observed that despite the effect of the decoherence program, the system remains coherent in the low-temperature limit. For a thorough investigation, we plotted the distillable coherence for the system in various temperatures and observed the same results. Also, we showed that in suitable time intervals between the measurements, the visibility of the interference fringes is nearly maximum at the low-temperature regime. Accordingly, we sought reasons for such behaviors.

In a search for the origin of the observed phenomenon, We investigated the entropy variations. We followed the changes in the entropy production in parallel with interference fringes at low and high temperatures. Interestingly, at low temperatures, the final entropy (when the decoherence process is completed) was far from the entropy of a thermal state, and, also we had interference fringes with high visibility. On the other hand, at high temperatures, the final entropy is identical to the entropy of a completely mixed state. Also, there are no interferences at the high-temperature regime. This observation leads us to understand the decoherence program as an entropy production process. More precisely, the variations of the entropy are the main reason for the system's behavior, and the system does not end up in a thermal state necessarily. In this regard, we defined the remained entropy as a new measure that allows us to control the evolution of the system in different situations.

Finally, after calculating the system equilibrium state, we studied the changes of internal energy for thermalization and a general quantum equilibration process. Since a thermal state is a passive state, we cannot extract ergotropy (maximum work extraction via a cyclic unitary evolution) from a thermal state. However, we showed that the extracted ergotropy from the final equilibrium state is not zero. Therefore, we cannot consider the system and thermal bath interaction a thermalization. The system equilibrates to a state that is not a thermal state. This study shows us that the nature of the quantum equilibration process is not in agreement with the thermalization process and is unknown yet.

\subsection*{Declarations}
\subsubsection*{Funding}
The author declare that no external funding was received for this study.

\subsubsection*{Conflict of Interests/Competing Interests}
The author declare no conflict of interests/competing interests.

\subsection*{Data Availability}
This work is purely mathematical and no external data is generated.

\end{document}